\documentclass[11pt]{article}
\usepackage{moriond,epsfig}

\usepackage{cite}
\usepackage{scalefnt}
\usepackage{verbatim}
\usepackage{amsmath}

\usepackage{psfrag}

\newcommand{\lt}{\left}
\newcommand{\rt}{\right}
\newcommand{\ds}{\displaystyle}
\newcommand{\tev}{\,\mbox{TeV}}
\newcommand{\gev}{\,\mbox{GeV}}
\newcommand{\mev}{\,\mbox{MeV}}

\newcommand{\Bbar}{\,\overline{\!B}}

\newcommand{\bbd}{$\mathrm{B_d}\!-\!\ov{\mathrm{B}}{}_\mathrm{d}\,$}

\newcommand{\bbms}{$\mathrm{B_s}\!-\!\ov{\mathrm{B}}{}_\mathrm{s}\,$\ mixing}
\newcommand{\bbmd}{$\mathrm{B_d}\!-\!\ov{\mathrm{B}}{}_\mathrm{d}\,$\ mixing}
\newcommand{\bbmq}{$\mathrm{B_q}\!-\!\ov{\mathrm{B}}{}_\mathrm{q}\,$\ mixing}
\newcommand{\kk}{$\mathrm{K}\!-\!\ov{\mathrm{K}}{}\,$}
\newcommand{\bb}{$\mathrm{B}\!-\!\ov{\mathrm{B}}{}\,$}
\newcommand{\bbm}{$\mathrm{B}\!-\!\ov{\mathrm{B}}{}\,$\ mixing}
\newcommand{\kkm}{$\mathrm{K}\!-\!\ov{\mathrm{K}}{}\,$\ mixing}

\newcommand{\dm}{\ensuremath{\Delta M}}
\newcommand{\dg}{\ensuremath{\Delta \Gamma}}

\newcommand{\eq}[1]{Eq.~(\ref{#1})}
\newcommand{\eqsand}[2]{Eqs.~(\ref{#1}) and (\ref{#2})}

\newcommand{\imag}{\mbox{Im}\,}

\newcommand{\bea}{\begin{eqnarray}}
\newcommand{\eea}{\end{eqnarray}}

\newcommand{\nn}{\nonumber \\}
\newcommand{\no}{\nonumber}
\newcommand{\ov}{\overline}
\newcommand{\epm}[2]{
 \raisebox{-0.5ex}{\shortstack[l]{$\scriptstyle+#1$\\$\scriptstyle-#2$}}}
\def\journalL#1#2#3#4#5{\journal{#1 #2}{#3}{#4}{#5}}
\def\journal#1#2#3#4{#1~{\bf #2}, #3 (#4)}

\def\PLB#1#2#3{\journal{Phys.\ Lett. B}{#1}{#2}{#3}}

\def\PRD#1#2#3{\journal{Phys.\ Rev. D}{#1}{#2}{#3}}

\newcommand{\arxiv}[1]{{arxiv:{#1}}}
\newcommand{\fig}[1]{Fig.~\ref{#1}}
\newcommand{\url}[1]{{\tt\small #1}}
\newcommand{\etal}{{\it et al.}}

\newcommand{\beqin}[1]{$ #1 $ }
\begin{document}
\vspace*{-1.7cm}
\title{Flavour physics, supersymmetry and grand
  unification\footnote{Talk at \emph{Rencontres de Moriond, EW
      Interactions and Unified Theories}, Mar 13th-20th, 2011, 
  La Thuile, Italy.}}

\author{Ulrich Nierste}

\address{Institut f{\"u}r Theoretische Teilchenphysik
      \\
      Karlsruhe Institute of Technology,
      Universit{\"a}t Karlsruhe
      \\
      Engesserstra\ss e 7, 76128 Karlsruhe, Germany}

\maketitle\abstracts{A global fit to quark flavour-physics data
  disfavours the Standard Model with 3.6 standard deviations and points
  towards new CP-violating physics in meson-antimeson mixing
  amplitudes. Tevatron data call for a new \bbms\ phase and new physics
  in \bbmd\ alleviates the tension on the unitarity triangle driven by
  $B(B\to\tau \nu)$.  In supersymmetric GUT models the large atmospheric
  neutrino mixing angle can influence $b\to s$ transitions. I present
  the results of a recent analysis in an SO(10) GUT model which
  accomodates the large \bbms\ phase while simultaneously obeying all
  other experimental constraints.}

\section{Introduction}
On May 17, 2010, \emph{The New York Times}\ wrote: 
\begin{center}
  \parbox{0.9\textwidth}{\emph{Physicists at the Fermi National
      Accelerator Laboratory are reporting that they have discovered a
      new clue that could help unravel one of the biggest mysteries of
      cosmology: why the universe is composed of matter and not its
      evil-twin opposite, antimatter.}}
\end{center}
This phrase was contained in an article featuring a measurement by the
D\O\ collaboration presented three days earlier by Guennadi Borissov in 
the talk  
\begin{center} 
\emph{Evidence for an anomalous like-sign dimuon charge asymmetry.} 
\end{center} 
D\O\ has studied the decays of pair-produced hadrons into final states
with muons \cite{dimuon_evidence_d0}. If, for example, a $(b,\ov b)$
pair hadronises into a $\Lambda_b$ baryon, a $B^+$ meson, and several
lighter hadrons, the semileptonic decays of $\Lambda_b$ and $B^+$ will
result in leptons of opposite charges. However, if the $b$ or $\ov b$
quark ends up in a neutral $B$ meson, \bb\ oscillations may lead to a
``wrong-sign'' muon charge: While a $B$ meson contains a $\ov b$ quark
decaying into a $\mu^+$, \bbm\ permits the process $B \to \ov B \to X
\mu^- \ov\nu_\mu$ resulting in a muon with negative charge.  The data
sample with like-sign dimuons is therefore enriched with events which
involve a mixed neutral meson. By further comparing the numbers of
$(\mu^-,\mu^-)$ and $(\mu^+,\mu^+)$ pairs in the final states D\O\ has
quantified the CP violation in \bbm\ for a data sample composed of $B_d$
and $B_s$ mesons. The central value of the measured CP asymmetry exceeds
the theory prediction \cite{ln} by a factor of 42 and the statistical
significance of the discrepancy is 3.2 standard
deviations.\footnote{After this conference D\O\ has updated the analysis
with a larger data sample and found a discrepancy of 3.9 standard
deviations with respect to the SM prediction \cite{dnew}.} A
new-physics interpretation of the measurement requires a large effect in
\bbms, because the precision measurements at the B factories limit the
size of a possible new CP phase in \bbmd. 

The Standard-Model (SM) predictions for the \bbm\ amplitudes involve
elements of the Cabibbo-Kobayashi-Maskawa (CKM) matrix, which are found
from global fits to many observables of flavour physics. CP-violating
quantities depend in a crucial way on the parameters $\ov \rho$ and $\ov
\eta$, which define the apex of the CKM unitarity triangle (UT) (see
Fig.~\ref{fig:box}).  $\ov \rho$ and $\ov \eta$ also govern the sizes of
$b\to u$ and $b\to d$ transitions. Among the quantities used in global
fits to the UT are the precisely measured \bbmd\ and \bbms\ oscillation
frequencies, the CP phase in \bbmd\ measured in the decay $B_d \to
J/\psi K_S$, and $\epsilon_K$, which quantifies CP violation in \kkm.
Several authors have noticed a tension in the Standard-Model (SM) fit of
the UT to data \cite{smtensions}. Meson-antimeson mixing amplitudes are
$\Delta F=2$ amplitudes, meaning that the flavour quantum number
$F=B,S,\ldots$ changes by two units. In a wide class of models beyond
the SM $\Delta F=2$ transitions receive larger new-physics corrections
than the $\Delta F=1$ decay amplitudes. The relation of the measured
quantities to the CKM elements will be altered if new physics affects
the $\Delta F=2$ amplitudes.  A proper theoretical assessment of the
quoted D\O\ measurement and of the tensions in the over-constrained CKM
matrix therefore calls for a global analysis which fits the elements of
the Cabibbo-Kobayashi-Maskawa (CKM) matrix simultaneously with complex
parameters quantifying new physics in \kk, \bbd, and \bbms. Such an
analysis has been performed in Ref.~\cite{lnckmf}.  In this talk I first
summarise the results of this analysis in Sec.~\ref{sec:ana}.
Subsequently, in Sec.~\ref{sec:susy} I interpret the results within 
a supersymmetric grand unified theory (GUT). In Sec.~\ref{sec:con} I
conclude.

\section{Anatomy of new physics in \bbm}\label{sec:ana}
In this section I present the essentials of the analysis in
Ref.~\cite{lnckmf}, which shows evidence for new physics (NP) in \bbm. While
I use the numerical ranges for experimental and theoretical input
quantities compiled in this reference, I use two simplifications in this
talk: First, for clarity of the presentation my quoted errors contain
statistical, systematic and theoretical errors added in quadrature. By
contrast, the original analysis in Ref.~\cite{lnckmf} has used the more
conservative Rfit procedure \cite{CKMfitter2}, which scans over
systematic and theoretical uncertainties. Second, whenever possible I
present simplified derivations of the tensions between experimental
results and SM predictions. In this way the main sources of the quoted
tensions become transparent.    

\begin{figure}
\hrule
\begin{center}
\includegraphics[width=0.34\textwidth]{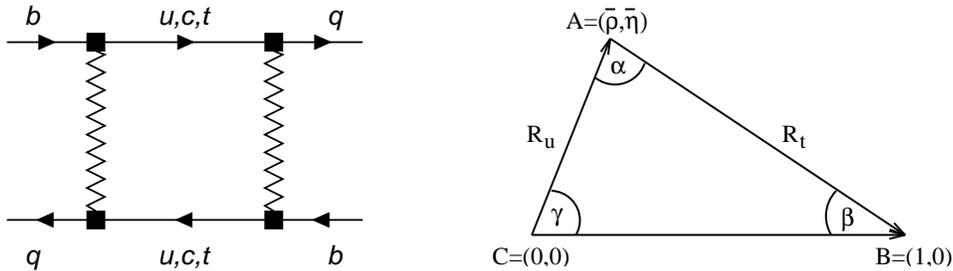}
~~~~~~
\includegraphics[width=0.38\textwidth,clip=true,]{figs/triangle2.ps}
\end{center}
\caption{Left: SM box diagram describing \bbmq, with $q=d$ or
  $s$. Right: Standard unitarity triangle.}\label{fig:box}
~\\[-7mm] \hrule
\end{figure}

Flavour-changing neutral current (FCNC) processes are known to be very
sensitive to NP. 
Schematically, any contribution to an FCNC $\Delta F=1$ decay amplitude
is proportional to $\delta_{\rm FCNC}/M^2$, where $\delta_{\rm FCNC}$ is
a small flavour-violating parameter and $M$ is some heavy mass scale.
In a SM diagram, $\delta_{\rm FCNC}$ is the product of two CKM elements
and the relevant scale $M$ is the $W$ boson mass entering the FCNC loop
diagrams. $\Delta F=2$ amplitudes, however, scale like $\delta_{\rm
  FCNC}^2/M^2$, for instance the $\Delta B=2$ box diagram of
\fig{fig:box} is proportional to $(V_{tb}V_{tq}^*)^2/M_W^2$. One
realises that $\Delta F=2$ amplitudes are more sensitive to NP than
$\Delta F=1$ transitions in a wide class of models: Whenever
$|\delta_{\rm FCNC}^{\rm NP}|>|\delta_{\rm FCNC}^{\rm SM}|$, which must
come with $M > M_W$ to keep the NP contribution smaller than the SM one,
the relative impact of NP on a $\Delta F=1$ transition is smaller by
factor of $|\delta_{\rm FCNC}^{\rm SM}|/|\delta_{\rm FCNC}^{\rm NP}|$
with respect to the $\Delta F=2$ case. In extensions of the SM with new
sources of flavour violation the case $|\delta_{\rm FCNC}^{\rm
  NP}|>|\delta_{\rm FCNC}^{\rm SM}|$ is the default situation, because
off-diagonal CKM elements are small.  Moreover, $\Delta F=1$ FCNC decays
hardly enter the global fit determining the CKM elements.  It is
therefore well-motivated to fit these elements in scenarios in which the
NP effects are confined to $\Delta F=2$ processes\cite{lnckmf}.

\subsection{The $|V_{ub}|$ puzzle}\label{sec:puz}
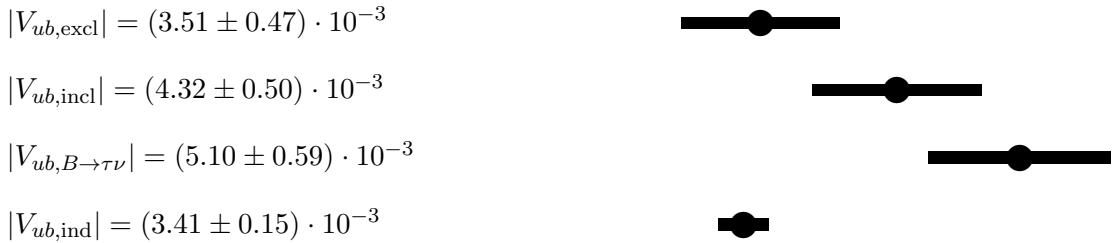
\begin{figure}
\vspace{-1.5cm}
\setlength{\unitlength}{0.14\textwidth} 
\hrule
\begin{picture}(7,1.5)
\linethickness{1.5mm}
\put(0,1.15){\makebox(3,0.3)[l]{
   \beqin{|V_{ub,\rm excl}|=(3.51 \pm 0.47)\cdot
      10^{-3}}}}
\put(4,1.3){\line(1,0){0.94}}
\put(4.47,1.3){\circle*{0.15}}
\put(0,0.75){\makebox(3,0.3)[l]{
   \beqin{|V_{ub,\rm incl}|=(4.32 \pm 0.50)\cdot
      10^{-3}}}}
\put(4.78,0.9){\line(1,0){1.0}}
\put(5.28,0.9){\circle*{0.15}}
\put(0,0.35){\makebox(3,0.3)[l]{%
   \beqin{|V_{ub,B\to \tau \nu}|=(5.10 \pm 0.59)\cdot
      10^{-3}}}}
\put(5.47,0.5){\line(1,0){1.08}}
\put(6.01,0.5){\circle*{0.15}}
\put(0,-0.05){\makebox(3,0.3)[l]{%
   \beqin{|V_{ub,\rm ind}|=(3.41 \pm 0.15)\cdot
      10^{-3}}}}
\put(4.22,0.1){\line(1,0){0.3}}
\put(4.37,0.1){\circle*{0.15}}
\end{picture}
\caption{Measurements of $|V_{ub}|$. The fourth value is 
indirectly obtained from the side $R_u$ of the UT.}\label{fig:vub}
~\\[-7mm] \hrule
\end{figure}

The CKM matrix\\[-8mm] 
\begin{eqnarray}
\hspace{-3mm} V_{\rm CKM} &=& \left(
\begin{array}{@{}ccc@{}}
\ds  V_{ud} &\ds   V_{us} &\ds   V_{ub} \\
\ds  V_{cd} &\ds   V_{cs} &\ds   V_{cb} \\
\ds  V_{td} &\ds  V_{ts} &\ds  V_{tb}
\end{array} \! \right) \no
\end{eqnarray}
is fixed by the measurements of 
\begin{equation}
|V_{us}|=0.2254 \pm 0.0013, \qquad\qquad\qquad  
|V_{cb}|=(40.9 \pm  0.7) \cdot 10^{-3},  \label{vusvcb}
\end{equation}
and the values of $\ov{\rho}$ and $\ov{\eta}$, which define the apex of
the unitarity triangle (UT) depicted in \fig{fig:box}: 
\begin{eqnarray}
  \ov{\rho} + i \ov{\eta} &\equiv&
  - \frac{{V_{ub}^*} V_{ud}} {{ V_{cb}^*} V_{cd}} \; 
  \equiv\;   R_u e^{i \gamma}
  \label{defut}
\end{eqnarray}
Currently $|V_{ub}|$ is measured in three ways, from i) the exclusive
decays $B\to \pi \ell\nu$, ii) the inclusive decays $B\to X \ell\nu$,
and iii) the leptonic decay $B^+ \to \tau^+ \nu_\tau$.  $B(B^+\to \tau^+
\nu_\tau)$ has been measured by both the BaBar and Belle collaboration,
each with two methods using either a leptonic or a hadronic tag
\cite{babe}, resulting in\footnote{After this conference the average
  $(1.64 \pm 0.34) \cdot 10^{-4}$ has been presented\cite{kowa}.}
\begin{eqnarray}
B^{\rm exp} (B^+\to \tau^+ \nu_\tau) &=&
  (1.68 \pm 0.31)  \cdot 10^{-4} .\no
\end{eqnarray}
The theory prediction involves the $B$ meson decay
constant $f_B$, which is calculated with the help of lattice QCD:
\begin{eqnarray}
 B(B^+\to \tau^+ \nu_\tau) & =& 1.13 \cdot 10^{-4} \cdot
         \lt( \frac{|V_{ub}|}{4\!\cdot\! 10^{-3}} \rt)^2 
         \lt( \frac{f_B}{200\, \mev} \rt)^2  \no
\end{eqnarray}
With \beqin{f_B= (191 \pm 13) \, \mev} one finds  
\begin{eqnarray}
 |V_{ub,B\to \tau \nu}| & =& 
   \lt[ 5.10 \pm \lt. 0.47\rt|_{\rm exp} \pm \lt. 0.35\rt|_{\rm f_B}
   \rt] \cdot 10^{-3} \nn 
  & =&                         
  \lt[ 5.10 \pm 0.59 \rt] \cdot 10^{-3} . \no
\end{eqnarray}
The measurement of $|V_{ub}|$ constrains the side $R_u$ of the UT,
because $|V_{ub}| \propto |V_{cb}| R_u$ (see \eq{defut}). However, in
the global fit to the UT the dominant constraint on $R_u$ stems from the
precise measurement of the mixing-induced CP asymmetry
$A_{\rm CP}^{\rm mix} (B_d \to J/\psi K_S)$. If the SM describes \bbmd\
correctly, this quantity determines the UT angle $\beta = 21.15^\circ
\pm 0.89^\circ$. Using further $\alpha=89^\circ
\epm{4.4^\circ}{4.2^\circ}$, we find $R_u= \sin\beta/\sin\alpha=0.361
\pm 0.015$ with a negligible impact of the error in $\alpha$. This value
is in excellent agreement with the result of the full global fit to the
UT. With our number for $R_u$ we can determine $|V_{ub}|$ indirectly
through \eq{defut}:
\begin{eqnarray}
 |V_{ub}|_{\rm ind} & =& 
  \lt( 3.41 \pm 0.15 \rt) \cdot 10^{-3} . \no 
\end{eqnarray}
The four determinations of $|V_{ub}|$ are shown in \fig{fig:vub}.
We observe no significant discrepancy between the individual direct
measurements of $|V_{ub}|$. However, there is a 2.9 $\sigma$ tension
between $B^+\to \tau^+ \nu$ and the indirect determination of $|V_{ub}|$ 
driven by $A_{\rm CP}^{\rm mix} (B_d \to J/\psi K_S)$.  

Several authors have studied NP contributions to $B^+ \to\tau^+ \nu$
\cite{chargedhiggs,right,new_Btau}. While a charged Higgs boson can
contribute to $B^+ \to\tau^+ \nu$, the contribution typically decreases
the branching fraction and therefore cannot solve the $|V_{ub}|$
puzzle. (A control channel for charged-Higgs effects is $B\to D\tau \nu$
\cite{B_Dtaunu}.) A more promising NP explanation of the $|V_{ub}|$
puzzle has been pointed out by Crivellin, who has observed that an
effective right-handed $W$ coupling $\ov b_R \gamma^\mu u_R W_\mu$ can
simultaneously shift $|V_{ub,\rm excl}|$ upwards and $|V_{ub,B\to \tau
  \nu}|$ downwards \cite{right}. The effect of a right-handed $W$
coupling on $|V_{ub,\rm ind}|$ is model-dependent. 

Since the direct determinations of $|V_{ub}|$ agree up to normal
statistical fluctuations, I argue that the simplest solution to the
$|V_{ub}|$ puzzle is NP in the \bbmd\ amplitude.  In the presence of a
new contribution $\phi_d^{\Delta}$ to the \bbmd\ phase the well-measured
$A_{\rm CP}^{\rm mix} (B_d \to J/\psi K_S)$ determines
$\sin(2\beta+\phi_d^{\Delta})$. With $\phi_d^{\Delta}<0$ the true value
of $\beta$ will be larger than $\beta = 21.15^\circ \pm 0.89^\circ$
inferred from the SM analysis. Since $\beta$ is also constrained by
other measurements, a global fit is required \cite{lnckmf}.

\subsection{New physics in \bbm}
\bbmq\ involves two hermitian $2\times 2$ matrices, the mass matrix
$M^q$ and the decay matrix $\Gamma^q$. The off-diagonal elements
$M_{12}^q$ and $\Gamma_{12}^q$ are calculated from the dispersive and
absorptive parts of the $\Bbar_q \to B_q$ transition amplitude,
respectively. In the SM $M_{12}^q$ is dominated by the box diagram in
\fig{fig:box} with internal top quarks, while $\Gamma_{12}^q$ stems from
box diagrams with only charm and up quarks on the internal lines.  The
SM expression for $M_{12}^q$ including NLO QCD corrections has been
calculated in Ref.~\cite{bjw}; the corresponding results for
$\Gamma_{12}^q$ have been obtained in Ref.~\cite{gamma,ln}. The
numerical predictions in Ref.~\cite{ln} have been recently updated with
present-day values of CKM elements, quark masses and hadronic parameters
in Refs.~\cite{lnckmf,ckmproc}. As a consequence of \bbmq, the mass
eigenstates $B_q^H$ and $B_q^L$ (with ``H'' and ``L'' denoting ``heavy''
and ``light'') found by diagonalising $M^q - i \Gamma^q/2$ are
linear combinations of $B_q$ and $\Bbar_q$. The mass and width differences
between   $B_q^H$ and $B_q^L$ are given by 
\begin{eqnarray}
\dm_q & = & M_H^q-M_L^q \; \simeq \; 2 |M_{12}^q| \, , \qquad\qquad
\dg_q \;=\; \Gamma_L^q-\Gamma_H^q \;\simeq \; 2 |\Gamma_{12}^q| \cos
\phi_q  \, .\no
\end{eqnarray}
The CP asymmetry in flavour-specific decays (such as $B_s \to X\ell^+
\nu_\ell$) reads 
\begin{eqnarray}
a_{\rm fs}^{q} &=& 
\frac{|\Gamma_{12}^q|}{|M_{12}^{q}|} \sin \phi_q \, \no
\end{eqnarray}
with the CP-violating phase\\[-7mm] 
\begin{eqnarray}
 \phi_q &\equiv& \arg \lt( - \frac{M_{12}^q}{\Gamma_{12}^q} \rt). \label{defphi}
\end{eqnarray}
The D\O\ measurement of the like-sign dimuon asymmetry
\cite{dimuon_evidence_d0,dnew} involves a sample which is almost evenly
composed of $B_d$ and $B_s$ mesons. The measured value is\footnote{The
  2011 value is $a_{\rm fs}= (-7.87\pm 1.72\pm 0.93)\cdot 10^{-3} $
  \cite{dnew}.}
\begin{eqnarray}
 a_{\rm fs} & =& (0.506 \pm 0.043) a_{\rm fs}^d +
   (0.494 \pm 0.043) a_{\rm fs}^s \nn
   & =&  ( - 9.57 \pm 2.51 \pm 1.46)\cdot 10^{-3}  
\label{dnumb}
\end{eqnarray}
Averaging with an older CDF measurement yields 
\begin{eqnarray}
a_{\rm fs}=\lt( -8.5
  \pm 2.8 \rt)\cdot 10^{-3}. \label{afsnum}
\end{eqnarray}
The numbers in \eqsand{dnumb}{afsnum} are $3.2 \sigma$ and $2.9 \sigma$
away from the SM prediction $a_{\rm fs}^{\rm SM}= \lt( -0.20\pm 0.03
\rt) \cdot 10^{-3}$ \cite{ckmproc}, respectively.

\begin{figure}
\hrule
\includegraphics[width=0.45\textwidth]
         {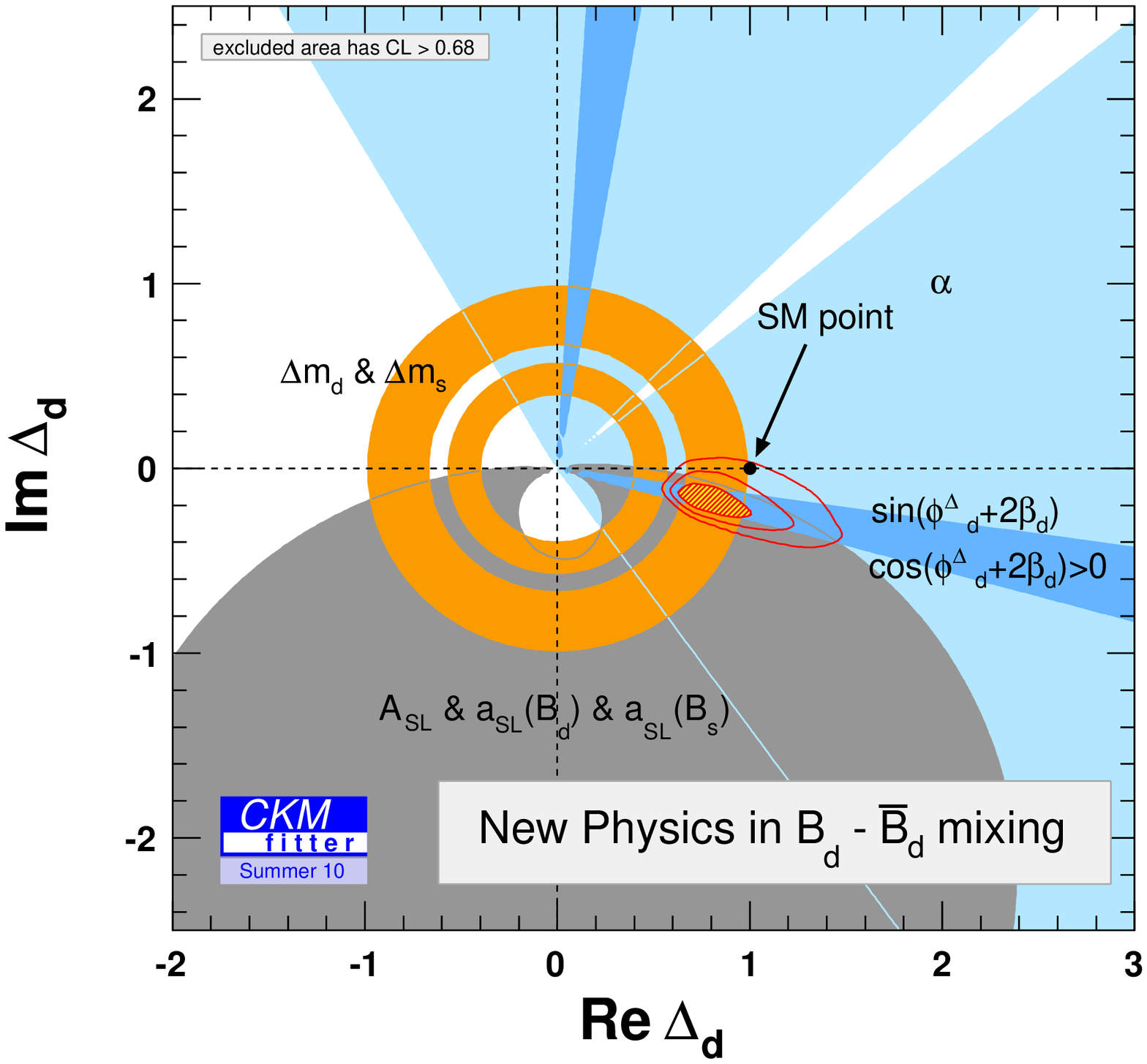}
~~~~~
\includegraphics[width=0.45\textwidth]
         {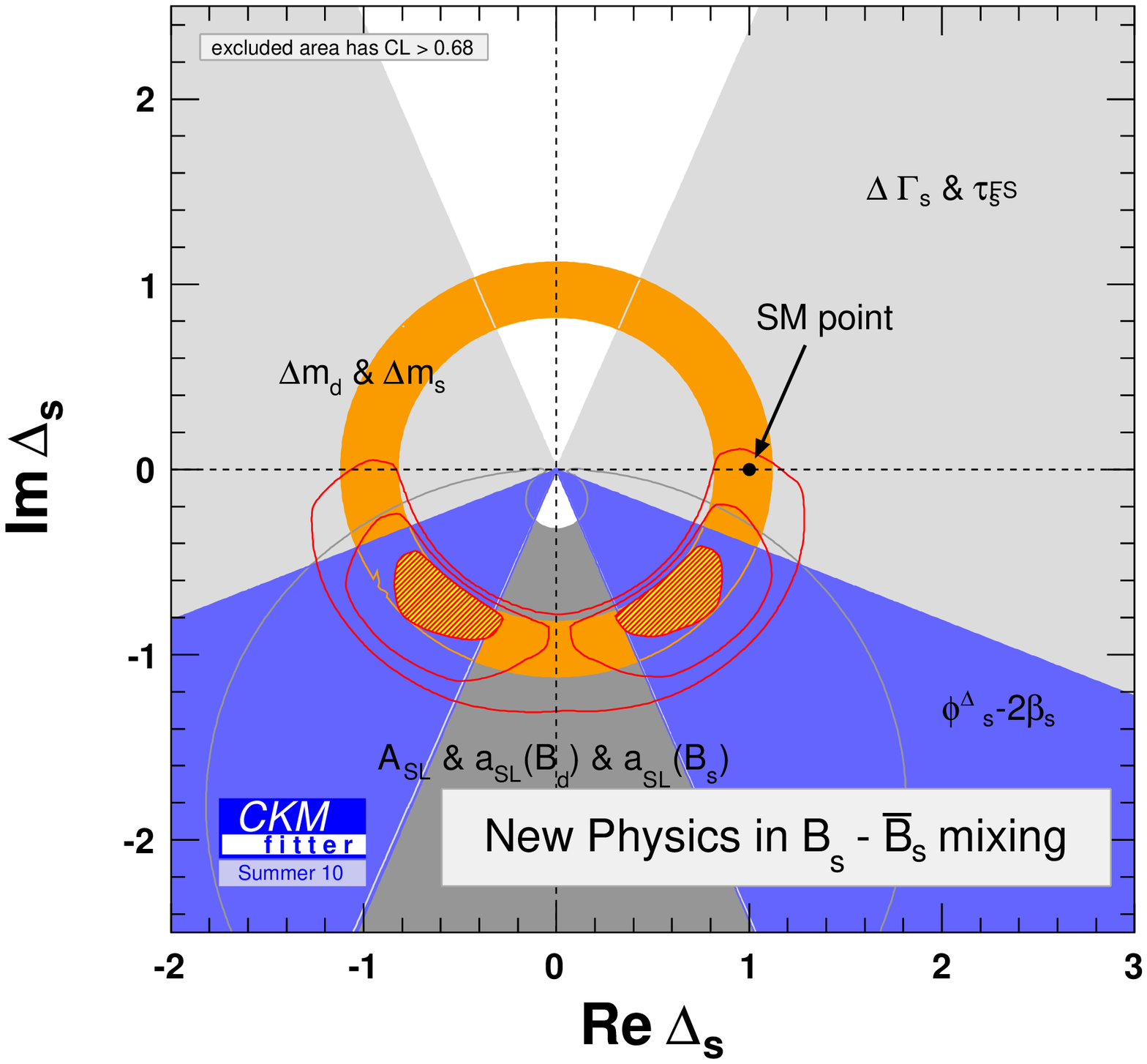}
         \caption{Allowed regions for $\Delta_d$ (left) and $\Delta_s$
           (right) from the global fit
           \cite{lnckmf}.}\label{fig:de}~\\[-7mm]\hrule
\end{figure}

$\Gamma_{12}^s$ originates from Cabibbo-favoured tree-level decays and
is insensitive to new physics.\footnote{Any NP competing with the
  tree-level $b\to s \ov c c$ decays constituting $\Gamma_{12}^s$ will
  alter the $b\to s \ov c c$ decay rates of all $b$-flavoured hadrons in
  conflict with the precisely measured charm content $n_c$ of $B$ decay
  final states and/or the semileptonic branching fraction
  \cite{lnckmf}.}  While $\Gamma_{12}^d$ involves some Cabibbo
suppression, it is nevertheless difficult to engineer a sizable
new-physics contribution to $\Gamma_{12}^d$ without running into
conflict with the plethora of measured exclusive $B$ decay branching
fractions. It is therefore safe to assume that NP contributions to
$\Gamma_{12}^{d,s}$ are irrelevant in view of today's experimental
errors.  In our analysis in Ref.~\cite{lnckmf} we have fitted the CKM
elements together with complex quantities parametrising new physics in
meson-antimeson mixing.  For \bbmq\ these parameters are defined as
\begin{eqnarray}
 \Delta_q & \equiv&
 \frac{ M_{12}^q}{M_{12}^{q,\rm SM}}, 
\qquad  \Delta_q \; \equiv \;  |\Delta_q| e^{i \phi^\Delta_q}, 
\qquad\mbox{with $q=d$ or $s$.}
\no
\end{eqnarray}
For \kkm\ one needs three such parameters. We have considered three
scenarios, with i) new physics with arbitrary flavour structure, ii)
minimally flavour-violating (MFV)\footnote{In MFV models all quark
  flavour violation is governed by the same CKM elements as in the SM.}
new physics with small bottom Yukawa coupling, and iii) MFV new physics
with large bottom Yukawa coupling.  These scenarios correspond to i)
$\Delta_d$, $\Delta_s$ complex and unrelated, ii)
$\Delta\equiv\Delta_d=\Delta_s$ real, and iii) $\Delta\equiv
\Delta_d=\Delta_s$ complex, respectively. In the first and third
scenario the constraint from $\epsilon_K$ is simply absent, because
\kkm\ is unrelated to \bbm, while in scenario ii) the \kkm\ NP
parameters can be expressed in terms of $\Delta$. In scenario i) we
obtain an excellent fit, the preferred regions in the complex
$\Delta_{d,s}$ planes are shown in \fig{fig:de}.
The point $\Delta_d=1$ is disfavoured by 2.7$\sigma$, and this
discrepancy is mainly driven by $B^+\to\tau \nu$ as discussed in
Sec.~\ref{sec:puz}. $\epsilon_K$ plays a minor role in our analysis
because of our conservative error estimate of the hadronic parameter
$B_K$. For a discussion of this issue see Soni's talk at this conference
\cite{sonitalk}.  $\Delta_s$ deviates from its SM value $\Delta_s=1$ by
2.7$\sigma$ as well, with $a_{\rm fs}$ as the main driver. Yet also the
CDF and D\O\ measurements of the CP phase in \bbms\ through $B_s \to
J/\psi \phi$ contribute here: Both measurements favour $\phi_s <0$, in
agreement with the conclusion drawn from $a_{\rm fs}$ in
\eq{afsnum}. The SM point $\Delta_d=\Delta_s=1$ is disfavoured with 3.6
standard deviations, establishing evidence of new physics.  Choosing a
different statistical test, $\imag\Delta_d=\imag\Delta_s=0 $ is even
disfavoured at a level of 3.8$\sigma$.

It is instructive to compare our best-fit result
\beqin{\phi_s^\Delta=(-52\epm{32}{25})^\circ} at 95\% CL with the
2010 Tevatron measurements. (I do not discuss the mirror solution
\beqin{\phi_s^\Delta=(-130\epm{28}{28})^\circ} in the third quadrant of
the complex $\Delta_s$ plane here.) The results of Ref.~\cite{cdfd02010}
read \beqin{\phi_s^\Delta=(-29 \epm{44}{49})^\circ} (CDF) and
\beqin{\phi_s^\Delta=(-44 \epm{59}{51})^\circ} (D\O) at \beqin{95\% \rm
  CL}. The naive average is \beqin{\phi_s^{\rm avg} = (-36 \pm
  35)^\circ} at 95\% CL. While the Tevatron measurements of
$\phi_s^\Delta$ alone contain only weak hints to new physics, they
perfectly agree with our best-fit value within normal statistical
fluctuations. If one discards $a_{\rm fs}$ in \eq{afsnum} altogether and
instead predicts it from the fit, one finds \beqin{a_{\rm fs}=\lt( -4.2
  \epm{2.9}{2.7}\rt)\cdot 10^{-3}} at \beqin{95\% \rm CL}, which is just
\beqin{1.5\sigma} away from the {D\O/CDF} average in \eq{afsnum}.  In
total a consistent picture of new physics in \bbm\ emerges, with a
normal upward statistical fluctuation of $a_{\rm fs}$ and a mild
downward fluctuation of the CDF value for \beqin{\phi_s^\Delta} from
$B_s \to J/\psi \phi$.

Scenario iii) also gives a reasonable fit to the data, but scenario ii) 
is as bad as the SM. This is bad news for the popular Constrained
Minimal Supersymmetric Standard Model (CMSSM) and its variant minimal
supergravity (mSUGRA), which are realisations of scenario ii).  

\section{Supersymmetry and grand unification}\label{sec:susy}
\begin{figure}
\hrule
\psfrag{amsq}{\scalefont{1.3}$\frac{a_1^d}{M_{\tilde
        q}}$\hspace{+0.3cm}} 
\psfrag{msq}{\hspace{-0.5cm}\scalefont{0.9}$M_{\tilde
      q}$[GeV]}
\psfrag{mg500argmu0tanb6}{}
\includegraphics[scale=.78, angle=0,clip=false]{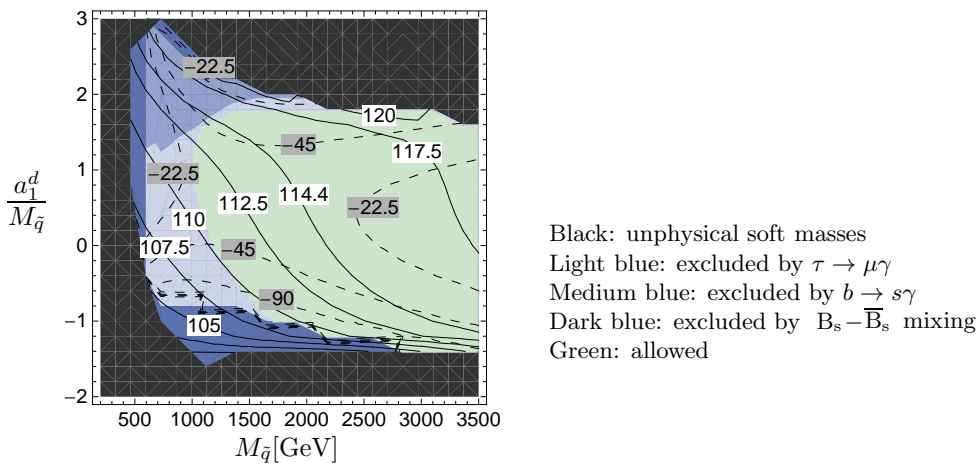}
~~~~\parbox[b]{0.4\textwidth}{\footnotesize
{Black:} unphysical soft masses \\
{Light blue:}  excluded by $ \tau\to \mu \gamma$\\
{Medium blue:} excluded by  $ b\to s \gamma$\\
{Dark blue:}  excluded by { \bbms}\\
{Green:} allowed\\[1cm]
}
\caption{Predictions of the CMM model for $m_{\widetilde
    g_3}=500\,\gev$, $\tan\beta =6$ and $\mu >0$. $M_{\widetilde q}$ is
  the (essentially degenerate) squark mass of the first two generations
  and $a_1^d$ is the trilinear supersymmetry-breaking term of the down
  squarks. The dashed lines with grey labels show the value of
  $\phi_s\simeq \phi_s^\Delta$ in degrees, the solid lines with white
  labels show the mass of the lightest neutral Higgs boson.  The black
  and blue regions are 
  excluded.~~~~~~~~~~~~~~~~~~~~}\label{fig:gut}~\\[-7mm]\hrule
\end{figure}

The MSSM has many new sources of flavour violation, all of which reside
in the supersymmetry-breaking sector. It is easy to get big effects in
\bbms, and the challenge is to suppress big effects elsewhere. MFV
variants of the MSSM cannot produce large effects in \bbms\ \cite{mfv}.
An attractive way to deviate from MFV in a controlled way (i.e.\ without
producing too large FCNC in observables agreeing with the SM) emerges if
one embeds the MSSM into a grand unified theory (GUT).  In a GUT quarks
and leptons reside in the same symmetry multiplets, which opens the
possibility of quark-flavour transitions driven by the leptonic mixing
matrix $U_{\rm PMNS}$ \cite{Moroi:2000tk,Chang:2002mq}. Consider 
  SU(5) multiplets: 
\begin{displaymath}
 {\bf \ov{5}_1} =
   \lt(\begin{array}{c} {d_R^c}\\ {d_R^c}\\ {d_R^c}\\
                e_L \\ -\nu_e \end{array} \rt), \qquad
 {\bf \ov{5}_2} =
   \lt(\begin{array}{c} {s_R^c}\\ {s_R^c}\\ { s_R^c}\\
                \mu_L \\ -\nu_\mu \end{array} \rt), \qquad
 {\bf \ov{5}_3} =
   \lt(\begin{array}{c} {b_R^c}\\ {b_R^c}\\ {b_R^c}\\
                \tau_L \\ -\nu_\tau \end{array} \rt). 
\end{displaymath}
If the observed large atmospheric neutrino mixing angle stems from a
rotation of $\bf \ov{5}_2$ and ${\bf \ov{5}_3}$, it will also affect the
$b_R$ and $s_R$ superfields. While rotations of quark fields in flavour
space are unphysical, this is not the case for the corresponding squark
fields $\widetilde b_R$ and $\widetilde s_R $ because of the
supersymmetry-breaking terms. The key ingredients of the idea of
Refs.~\cite{Moroi:2000tk,Chang:2002mq} is the following: In a weak basis
with diagonal up-type Yukawa matrix the down-type Yukawa matrix
$\mathsf{Y}_d$ is diagonalised as $\mathsf{Y}_d= V_{\rm CKM}^*
\mbox{diag}\, (y_d,y_s,y_b) U_{\rm PMNS} $. In this basis the
right-handed down-squark mass matrix has the form $\mathsf{m}^2_{\tilde
  d} = \mbox{diag}\, ( m^2_{\tilde d}, \, m^2_{\tilde d}, \, m^2_{\tilde
  d} - \Delta_{\tilde d} ) $ with a calculable real parameter $
\Delta_{\tilde d}$ generated by top-Yukawa renormalisation
group effects.  Rotating now $ \mathsf{Y}_d$ to diagonal form puts the
large atmospheric neutrino mixing angle into $\mathsf{m}^2_{\tilde d}$:
\begin{align}
   U_{\rm PMNS}^\dagger \, \mathsf{m}^2_{\tilde d}\, U_{\rm PMNS} &=
  \begin{pmatrix}
    m^2_{\tilde d} & 0 & 0 \\ 
    0 & m^2_{\tilde d} - \frac{1}{2}\,
    \Delta_{\tilde d} & - \frac{1}{2}\, \Delta_{\tilde d}\,
   e^{i\xi} \\ 0 & - \frac{1}{2}\, \Delta_{\tilde d}\,    
   e^{-i\xi} & m^2_{\tilde d} - \frac{1}{2}\, \Delta_{\tilde d}
  \end{pmatrix} \no
\end{align}
As a result we find large new transitions between right-handed
$\widetilde b$ and $\widetilde s$ squarks while keeping all other quark
FCNC transitions MFV-like. Moreover, the CP phase $\xi$ affects \bbms!
The GUT boundary conditions further connect $b_R \to s_R$ with $\tau_L
\to \mu_L$ transitions, so that \bbms\ is correlated with $\tau \to \mu
\gamma $.  The CMM model realises this idea using the GUT symmetry
breaking chain SO(10)$\to$ SU(5) $\to$
SU(3)$\times$SU(2)$_L\times$U(1)$_Y$.  In Ref.~\cite{gjkmnsw} we have
performed a global analysis of the CMM model, considering flavour
physics data, vacuum stability bounds and the lower bounds on sparticle
masses and the mass of the lightest Higgs boson. All MSSM parameters
involved depend on just seven CMM-model parameters.  We find that we can
accommodate a large \bbms\ phase while simultaneously obeying all other
experimental constraints (see \fig{fig:gut}).
In the CMM model eight of the twelve squark masses are essentially degenerate 
and are typically larger than 1$\,\tev$, as can be seen from
\fig{fig:gut}. Finally, corrections to the Yukawa  couplings from dimension-5 
terms can leak some of the CMM contribution in \bbms\ to \bbd\ and
\kkm\ and alleviate the tension in the fit to the UT
\cite{Trine:2009ns}.   

\section{Conclusions}\label{sec:con}
Precision data of flavour physics put the Standard Model under
pressure. The global analysis of Ref.~\cite{lnckmf} disfavours the SM at
a level of 3.6$\sigma$ and reveals a consistent picture of new
CP-violating physics in meson-antimeson mixing. The data cannot be
accommodated in the popular CMSSM and mSUGRA scenarios. However, the
large CP phase in \bbms\ can naturally be explained in GUT models which
link the large atmospheric neutrino mixing angle to novel $b\to s$
transitions \cite{Moroi:2000tk,Chang:2002mq}. Our recent quantitative
analysis, which relates FCNC observables, the Higgs mass and other
theoretical and experimental constraints to just seven parameters, has
found that this idea is indeed viable and permits large effects in
\bbms\ \cite{gjkmnsw}. 

\section*{Acknowledgements}
I thank the organisers for inviting me to this conference. I appreciate
the enjoyable collaborations with A.~Lenz, J.~Charles,
S.~Descotes-Genon, A.~Jantsch, C.~Kaufhold, H.~Lacker, S.~Monteil,
V.~Niess, S.~T'Jampens, J.~Girrbach, S.~J\"ager, M.~Knopf, W.~Martens,
C.~Scherrer and S.~Wiesenfeldt on the presented results. I thank
A.~Crivellin for proofreading the manuscript.  My work was supported by
DFG through grant No.~NI 1105/1-1, project C6 of the CRC--TR 9 and by
BMBF through grant no.~05H09VKF.

\end{document}